\documentclass[a4paper,12pt]{article}
\pdfoutput=1 
\usepackage{xcolor}
\usepackage{a4wide}
\usepackage[super]{cite}
\usepackage{mathpazo}
\usepackage{graphicx}
\usepackage[footnotesize, labelfont=bf,it]{caption}
\usepackage{amsmath}
\usepackage{algorithm}
\usepackage[noend]{algpseudocode}
\usepackage{hyperref}
\usepackage{lineno}
\usepackage{calrsfs}
\usepackage{xr}

\DeclareMathAlphabet{\pazocal}{OMS}{zplm}{m}{n}

\newcommand{\nc}{\newcommand}
\nc{\mb}{\mathbb}
\nc{\N}{\mb N}
\nc{\R}{\mb R}
\DeclareMathOperator*{\argmax}{argmax}

\author{Rohit Sahasrabuddhe$^{1}$, Renaud Lambiotte$^{2,3}$ \& Laura Alessandretti$^{4,*}$}
\title{From centre to centres: \\ polycentric structures in individual mobility}

\begin{document}

\maketitle

$^1$ Indian Institute of Science Education and Research, Pune, India\\
$^2$ Mathematical Institute, University of Oxford, United Kingdom \\
$^3$ Turing Institute, London, United Kingdom \\
$^4$ Technical University of Denmark, Anker Engelundsvej 1, Kgs Lyngby, DK-2800, Denmark\\

$^*$ To whom correspondence should be addressed, lauale@dtu.dk.\\

\pagebreak

\section*{Abstract}
\textbf{%
The availability of large-scale datasets collected via mobile phones has opened up opportunities to study human mobility at an individual level. The granular nature of these datasets calls for the design of summary statistics that can be used to describe succinctly mobility patterns.
In this work, we show that the radius of gyration, a popular summary statistic to quantify the extent of an individual's whereabouts, suffers from a sensitivity to outliers, and is incapable of capturing mobility organised around multiple centres. We propose a natural generalisation of the radius of gyration to a polycentric setting, as well as a novel metric to assess the quality of its description. With these notions, we propose a method to identify the centres in an individual's mobility and apply it to two large mobility datasets with socio-demographic features, showing that a polycentric description can capture features that a monocentric model is incapable of. 
}


The availability of individual-level mobility data collected from mobile phones has taken the study of human travel to a new direction, leading to fundamental discoveries on the mechanisms driving spatial behaviour \cite{schlapfer2021universal,alessandretti2020scales,gonzalez2008understanding,song2010limits}, and enhancing our ability to address societal challenge, from forecasting the spread of epidemics \cite{pepe2020covid,hunter2021effect,chang2021mobility}, to designing sustainable transport solutions \cite{song2016deeptransport,dabiri2018inferring,zhu2018big}. While opening a wide range of new opportunities, the need to store, process and anonymise large scale individual trajectory datasets is also raising new challenges, and calling for the development of novel analytical and computational solutions \cite{barbosa2018human,de2018privacy,lazer2021meaningful}.

One important issue relates to designing summary statistics that, given an individual sequence of raw spatial coordinates, capture relevant aspects of spatial behaviour, such as the regularity and predictability of a person's movements \cite{song2010limits,lu2013approaching}, the evolution of visitation patterns over long time-scales \cite{alessandretti2018evidence} or the interplay between exploration and exploitation of locations \cite{alessandretti2018evidence}. It is critical to identify measures that provide concise and accurate descriptions of these aspects, but still allow to capture the heterogeneity of behaviours across individuals. 

In this work, we address this challenge focusing on one of the most widely studied aspects of spatial behaviour, the area characterising individuals' whereabouts, e.g. the region within which people move during the course of their daily activities. Following the geographical literature, we refer to this concept as to the \emph{activity space} \cite{zenk2011activity,wong2011measuring,sherman2005suite}. Note that similar ideas are investigated in animal ecology, where researchers refer to the \emph{home-range} as to the region that encompasses the resources an animal requires to survive and reproduce \cite{hayne1949calculation,laver2008critical,powell2012home}.

A number of approaches were proposed to operationalise the concept of \emph{activity space} \cite{sherman2005suite,powell2012home}. In recent years, the \emph{radius of gyration} approach inspired by the physics of rigid bodies has gained widespread popularity in the Human Mobility literature, because of its simple formulation and scalable computation time \cite{gonzalez2008understanding,song2010modelling}. This model, which we call the \textit{monocentric model}, proposes that, for each individual, the activity space can be summarised as a circle extending on the earth surface. The centre of the circle, the so-called \emph{centre of mass}, and its radius, the so-called \emph{radius of gyration}, are determined by the positions of the locations visited by the individual and their importance in terms of visitation probability. 
Following its introduction in the pioneer article by Gonzalez et al \cite{gonzalez2008understanding}, the monocentric model has been used for many applications, including characterising regularities across individuals \cite{hawelka2014geo,pappalardo2015returners,lu2013approaching,song2010limits,jurdak2015understanding,cheng2011exploring}, modelling the spread of epidemics \cite{pepe2020covid,merler2010role}, studying patterns of crime activity \cite{bernasco2019adolescent,curtis2021national}, and movements during natural disasters \cite{wang2014quantifying,wang2016patterns}.

\begin{figure}
    \centering
    \includegraphics[width=\textwidth]{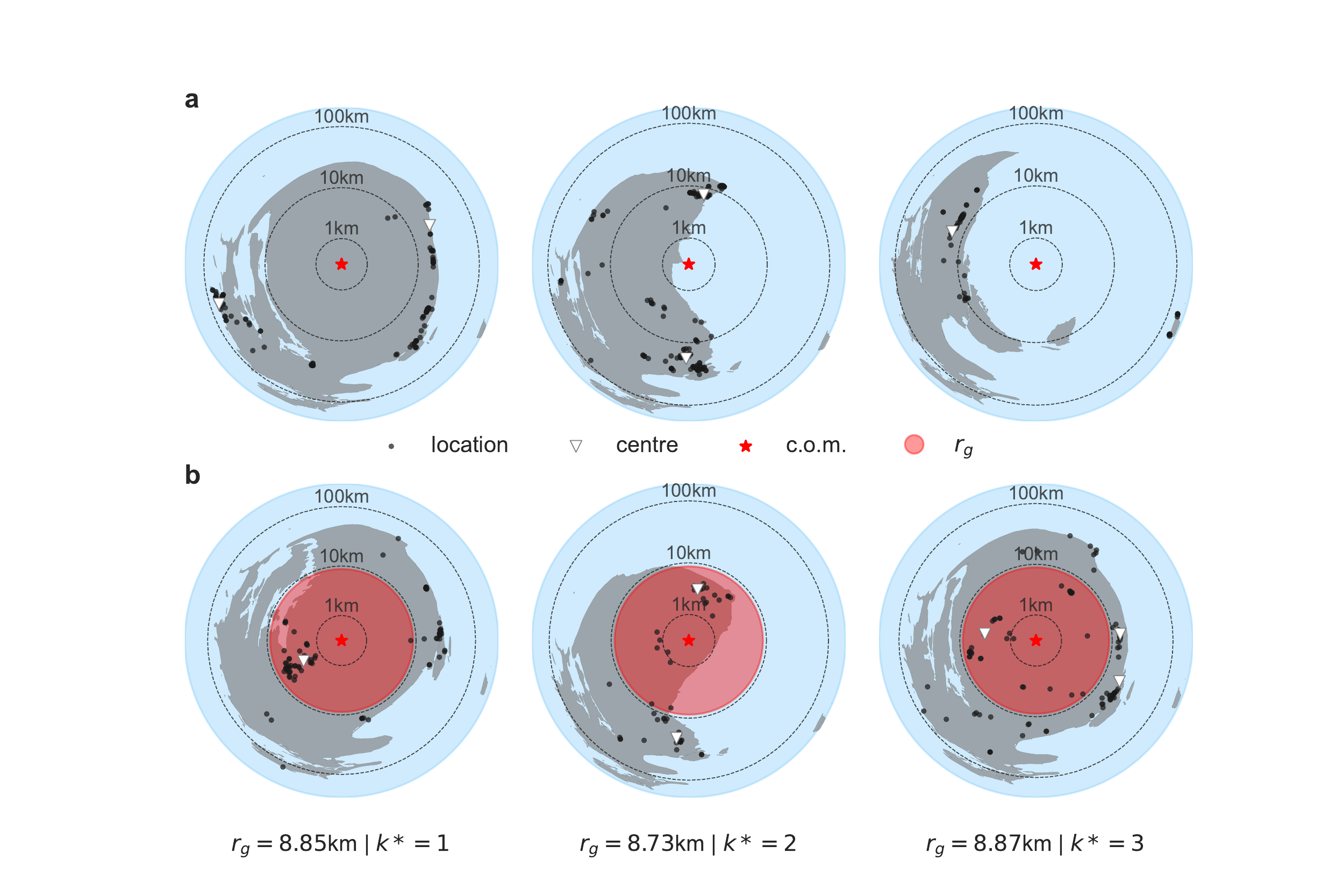}
    \caption{\textbf{Inadequacy of the monocentric description of activity spaces.} Each map shows an individual's activity space displayed using a log-azimuthal projection \cite{snyder1987magnifying}. Individuals (all from dataset D1, see data description in the \emph{Methods} section) are based in the Zealand region of Denmark. \textbf{(a) Position of the centre of mass.} In all examples (left to right), the centre of mass (red star) lies far from any of the visited locations (black dots). Our method identifies multiple centres (white triangles) closer to individual activities. \textbf{(b) Extent of the activity space.} The examples (left to right) show the activity spaces for three individuals with similar radius of gyration (red circles), but different number of centres, according to our method (white triangles). The radius of gyration is unable to capture the wide differences in the extent of whereabouts across the three individuals.}
    \label{fig:inadequate_panel}
\end{figure}

While being simple and concise, the description of individual whereabouts as a circle has recently been challenged by empirical findings revealing that mobility is organised around multiple centres. Individual daily movements are anchored around several focal points, including home, work, and relevant points of interests such as schools, family homes etc \cite{bagrow2012mesoscopic,schneider2013unravelling} It has further been shown that the organisation of infrastructure, from cities to locations within cities, is polycentric \cite{louf2013modeling} and hierarchical \cite{bassolas2019hierarchical}, and this structure affects human travel \cite{alessandretti2020scales}. Due to its monocentric nature, the \emph{radius of gyration} approach for describing activity spaces may, in some cases, fail to accurately capture spatial behaviour. We identify at least two anomalies: (i) the \emph{position} of the centre of mass may be far from a person's actual whereabouts, because, when locations are organised around multiple focal points, the centre of mass may be located in an unfamiliar area (see Fig.~\ref{fig:inadequate_panel}a); and (ii) comparing the \emph{extent} of the activity space, measured by the radius of gyration, for individuals with polycentric mobility may be misleading. For example, individuals with the same radius of gyration but different number of centres have widely different spatial behaviour (see Fig.~\ref{fig:inadequate_panel}b for an illustration). 

In this work, we show that a simple generalisation of the radius of gyration description can capture the polycentricity of human mobility. We propose a new metric, the \emph{cover score}, that allows us to assess the quality of a mobility description, by quantifying the interplay between its extent and the time spent within it. 
In doing so, we distinguish between two different spatial scales in human mobility, one associated to the distance between centres, and one to the distance around the centres.
We propose a new algorithm to compute polycentric activity spaces, and we validate it against synthetic data. We define the description arising from our method as the \emph{flexicentric model}, because it accommodates both monocentric and polycentric activity spaces. We further validate our approach using real-world data: first, we formulate a simple time allocation model that, given an activity space, describes where individuals spend time; then, we show that, under this simple time allocation model, the flexicentric description outperforms the monocentric model at the task of predicting unseen data without overfitting. Finally, we use the proposed method to extensively characterise activity spaces across two large-scale datasets describing the GPS trajectories of $\sim{50,000}$ individuals (see the \emph{Methods} section), in relation to socio-demographic attributes.

\section*{Results} \label{sec:results}

\textbf{From monocentric to polycentric description of individual whereabouts.} 
 Consider the set $L = \{l_1,...,l_i,...l_N\}$ of $N$ locations visited by an individual within a certain period of time, where each visited location $l_i$ is characterised by its geographical position $\mathbf{x_i}$, and its weight $w_i$, defined as the fraction of time spent in location $l_i$ within the period considered, such that $\sum_{i=1}^N w_i = 1$. In the monocentric model, one summarises the whereabouts of this individual as a circle centred at the so-called \emph{centre of mass}: 
\begin{equation}
    \mathbf{c.o.m.} = \sum_{i = 1}^N w_i{\mathbf{x_i}},
    \label{eq:com}
\end{equation}

with radius equal to the so-called \emph{radius of gyration}:

\begin{equation}
    r_g = \sqrt{\sum_{i = 1}^N w_i \; (\mathbf{x_i} - \mathbf{c.o.m.})^2}.
    \label{eq:rog}
\end{equation}
 The origin of this quantity is rooted in the physics of rigid bodies. In their pioneering article\cite{gonzalez2008understanding}, the authors drew an analogy between a 2-dimensional rigid body and the set of individual locations $L$ described above. In classical mechanics, a rigid body can be described as a set of point masses, where each point is characterised by its position and mass \cite{goldstein2002classical}. Similarly, in the study of mobility, each location $l_i$ is characterised by its position $\mathbf{x_i}$ and its weight $w_i$. Further, the \emph{moment of inertia} of a rigid body quantifies the spread of its mass around a given axis, and physically relates to how difficult it is to rotate the body about the axis. For a 2-dimensional body, the moment of inertia can be defined with respect to a centre $\mathbf{c}$, where the axis of rotation is perpendicular to the plane of the body, and intersects it at $\mathbf{c}$. Using our notation, the moment of inertia is defined as:
\begin{equation}
    I(\mathbf{c}) = \sum_{i = 1}^N w_i\; (\mathbf{x_i} - \mathbf{c})^2.
    \label{eq:inertia}
\end{equation}
 It can be proven \cite{goldstein2002classical} that, choosing $\mathbf{c}=\sum_{i = 1}^N w_i \mathbf{x_i}$ (the \emph{centre of mass}) minimises the moment of inertia. The corresponding 'characteristic distance' is the so-called \emph{radius of gyration}, $r_g = \sqrt{\frac{I(\mathbf{c.o.m.})}{\sum_{i = 1}^N w_i}}$ (see Eq.~\ref{eq:rog}), and describes a circle around the centre of mass, such that if the entire mass were concentrated upon it, the mechanical properties of the body would be preserved. 
Note that the radius of gyration  also has a simple statistical interpretation, as it is the standard deviation of the data points when $w_i$ is interpreted as their frequency in the Euclidean space.

Building on this analogy, the quantities described in Eq.~\ref{eq:com} and Eq.~\ref{eq:rog} implicitly model an individual's activity space as a circle, and have been used in a variety of contexts \cite{hawelka2014geo,pappalardo2015returners,lu2013approaching,song2010limits,jurdak2015understanding,cheng2011exploring,pepe2020covid,merler2010role,bernasco2019adolescent,curtis2021national,wang2014quantifying,wang2016patterns}. Yet, recent observational studies have documented polycentricity in human mobility, which an approach identifying only one centre and one spatial scale fails to capture \cite{louf2013modeling,bassolas2019hierarchical,alessandretti2020scales,bagrow2012mesoscopic}. 
To address this important limitation, the monocentric model can be extended to account for the polycentric nature of individual activity spaces. Specifically, we propose that an individual's activity space can be described as a set of $k$ ($k \geq 1$) circles, with centres $\{ \mathbf{c_1}, \mathbf{c_2}, \dots \mathbf{c_k} \}$. Under this flexicentric model, the \textit{moment of inertia} (see Eq.~\ref{eq:inertia}) can be generalised as:

\begin{equation}
 I(\{ \mathbf{c}_1, \mathbf{c}_2, \dots \mathbf{c}_k \}) = \sqrt{\sum_{i = 1}^N w_i \, \left(\min_{\alpha \in \{1,2,\dots k \}} |\mathbf{x_i} - \mathbf{ c}_\alpha |\right)^2} 
  \label{eq:new_inertia}
\end{equation}

Using this new definition, the inertia quantifies the weighted average squared distance of each location from its closest centre. Clearly, setting $k=1$ reduces the formula above to the monocentric description of inertia (see Eq.~\ref{eq:inertia}). Analogous to the monocentric model, we can choose the best set of centres as those which minimise the generalised inertia (Eq.~\ref{eq:new_inertia}).

\textbf{Identifying centres.} We note that the quantity defined in Eq.~\ref{eq:new_inertia} corresponds precisely to the loss function of the k-means clustering algorithm \cite{likas2003global}. The k-means algorithm aims to partition a set of weighted points into $k$ clusters in such a way that each point belongs to the cluster with the nearest weighted mean (see also the \emph{Methods} section). In our case, the set of points corresponds to the set of locations $L$ defined in the previous subsection. Given this correspondence, we could simply use the k-means algorithm to identify the centres minimising the inertia defined in Eq.~\ref{eq:new_inertia}.

However, it is known that the solution of k-means clustering is sensitive to outliers \cite{hautamaki2005improving}, which are quite common in our case due to the broad distribution of the location weights \cite{gonzalez2008understanding,song2010limits}. Hence, we suggest a modified version of the k-means algorithm, the trimmed-k-means (t-k-means), which accounts for the presence of outliers by trimming out the furthest points from the identified centres. Note that similar improvements of the k-means algorithm have been suggested by existing literature \cite{cuesta1997trimmed}\cite{pelleg2000xmeans}\cite{kmedoids}(see the description of the algorithm in the \emph{Methods} section). While accounting for outliers, the t-k-means algorithm does not minimise the inertia in Eq.~\ref{eq:new_inertia}.

\textbf{A novel metric to assess the quality of a model.} The t-k-means algorithm provides us a way to find centres for a given value of $k$, but what is the optimal value of $k$? This question is notoriously challenging and generally arises in clustering problems \cite{newman2016estimating,hamerly2004learning}.
In this section, we introduce the \emph{cover score} - a new metric specifically designed for assessing the quality of a description of an individual's activity space. 

Having identified the centres for a given choice of $k$, we describe individual's whereabouts as a set of $k$ circles around these points. For simplicity, all the circles have the same radius $r_k$. Here, we note that the choice of $r_k$ is arbitrary, but should allow for comparisons across choices of $k$. Thus, we choose $r_k$ such that the sum of the areas of the $k$ circles is equal to the area of the circle described by the radius of gyration. By setting $\pi k r_k^2 = \pi r_g^2$, we find $r_k=\frac{r_g}{\sqrt{k}}$. 

\begin{figure}
    \centering
    \includegraphics[width=\textwidth]{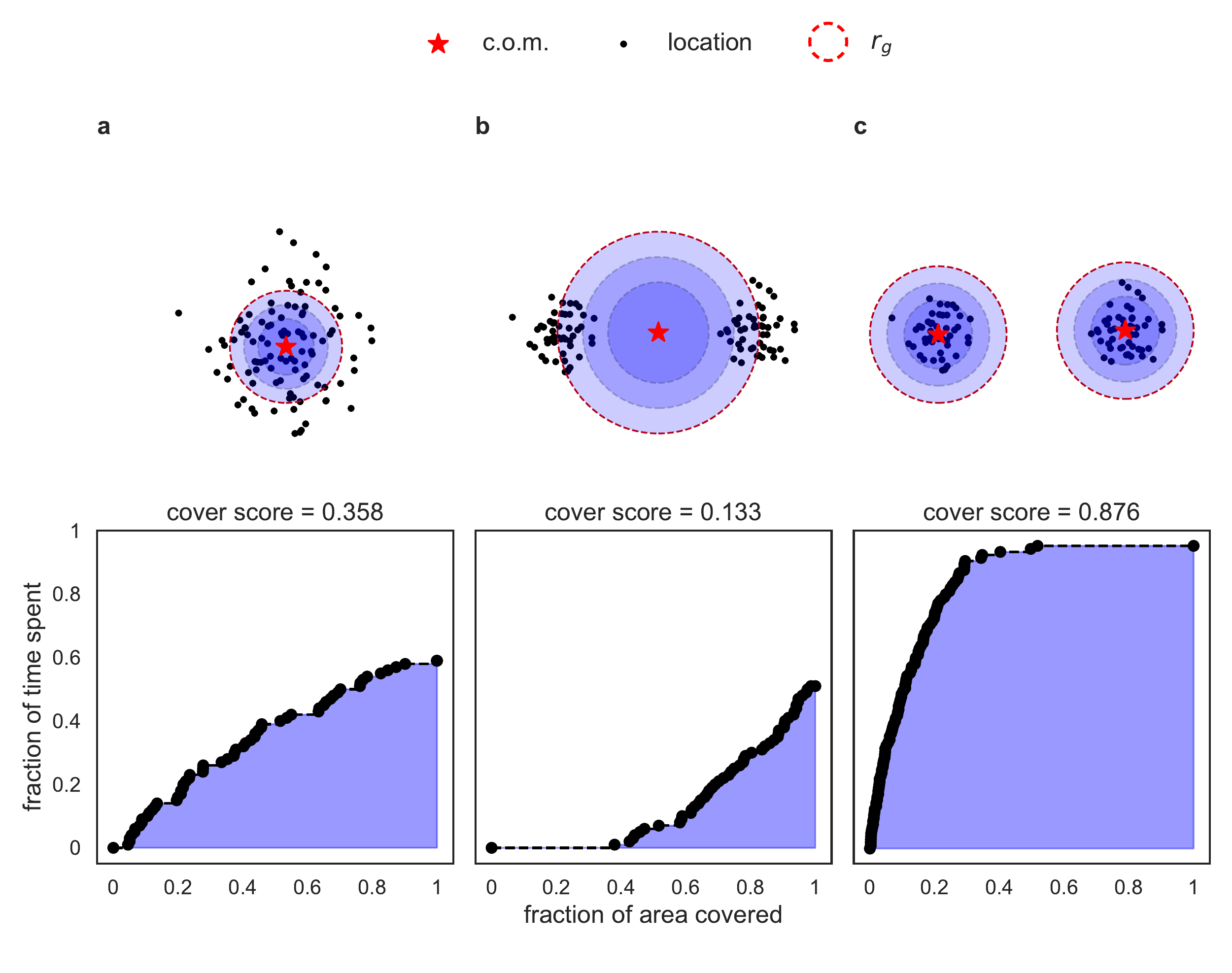}
    \caption{\textbf{Illustration of the cover score on synthetic data.} In the top panels we show the distribution of locations in space (black dots), the centres (star markers), and the corresponding circles with radius increasing from $0$ (blue circles, from dark to lighter) up to the radius of gyration (dashed red line). In the bottom panels, we show the curve describing the fraction of time spent inside the circles, $T(r)$ as a function of the fraction of the total area within the circles, $A(r)$. The area under the curve (coloured in blue) corresponds to the cover score. Results are shown in \textbf{(a)} for 100 locations sampled from a 2-dimensional Gaussian distribution with $\sigma^2=4$ along both axes; and in \textbf{(b)} and \textbf{(c)} for 100 locations sampled (50 each) from two Gaussian distributions with $\sigma^2=4$ along both axes, centred at $(0,0)$ and $(20,0)$. A monocentric representation is presented in \textbf{(b)}, as well as a polycentric ($k=2$) one in \textbf{(c)}. For simplicity, all locations have the same weight.}
    \label{fig:auc_schematic_all}
\end{figure}

The cover score quantifies the interplay between the area of the circles and the time spent within the space defined by the circles, as we increase their size. It is computed for a given set of centres as follows:
\begin{itemize}
\item Consider a set $C(r)$ of $k$ circles located at the centres, where each circle has radius $r$. Set $r=0$.
\item Increase the radius $r$ of the circles at the same rate, from $r=0$ up to $r=r_k$. As the circles grow, compute: 
\begin{itemize}
\item[(i)]{the fraction of time spent in locations located inside any of the circles $T(r)=\sum_{i \in \{j \, |\, l_j \in C(r)\}} w_i$, where $w_i$ are the weights of the locations located within any of the circles in $C(r)$ (in our notation, $l_i \in C(r)$) }
\item[(ii)]{the ratio $A(r)= \frac{r^2}{r_g^2}$ between the total area covered by the set of circles $C(r)$, and the maximum area, obtained when the circles have radius $r=r_k$.}
\end{itemize}

\item Compute the \textit{cover score} as the area under the curve $T(r)$ vs $A(r)$, i.e. the definite integral of the curve describing the growth of time spent within the activity space as its size increases (see Fig.~\ref{fig:auc_schematic_all})

\end{itemize}

Intuitively, given an activity space with $k$ centres, the cover score is high when activity is concentrated close to the centres, and low if the activity is concentrated far from the centres. In Fig.~\ref{fig:auc_schematic_all}, we illustrate how the cover score is computed using synthetic data. Considering a set of $100$ locations sampled from a 2-dimensional Gaussian distribution, we find that the monocentric description provides a good representation of the data, with a cover score of $0.358$ (Fig.~\ref{fig:auc_schematic_all}a). Instead, considering locations generated from two Gaussian distributions, we find that the monocentric model is not suitable (Fig.~\ref{fig:auc_schematic_all}b, top panel), as we achieve a low cover score of $0.133$ (Fig.~\ref{fig:auc_schematic_all}b, bottom panel), while using two centres to describe the same data (Fig.~\ref{fig:auc_schematic_all}c, top panel), results in a much higher score of $0.876$ (Fig.~\ref{fig:auc_schematic_all}c, bottom panel).

\begin{figure}
    \centering
    \includegraphics[width=\textwidth]{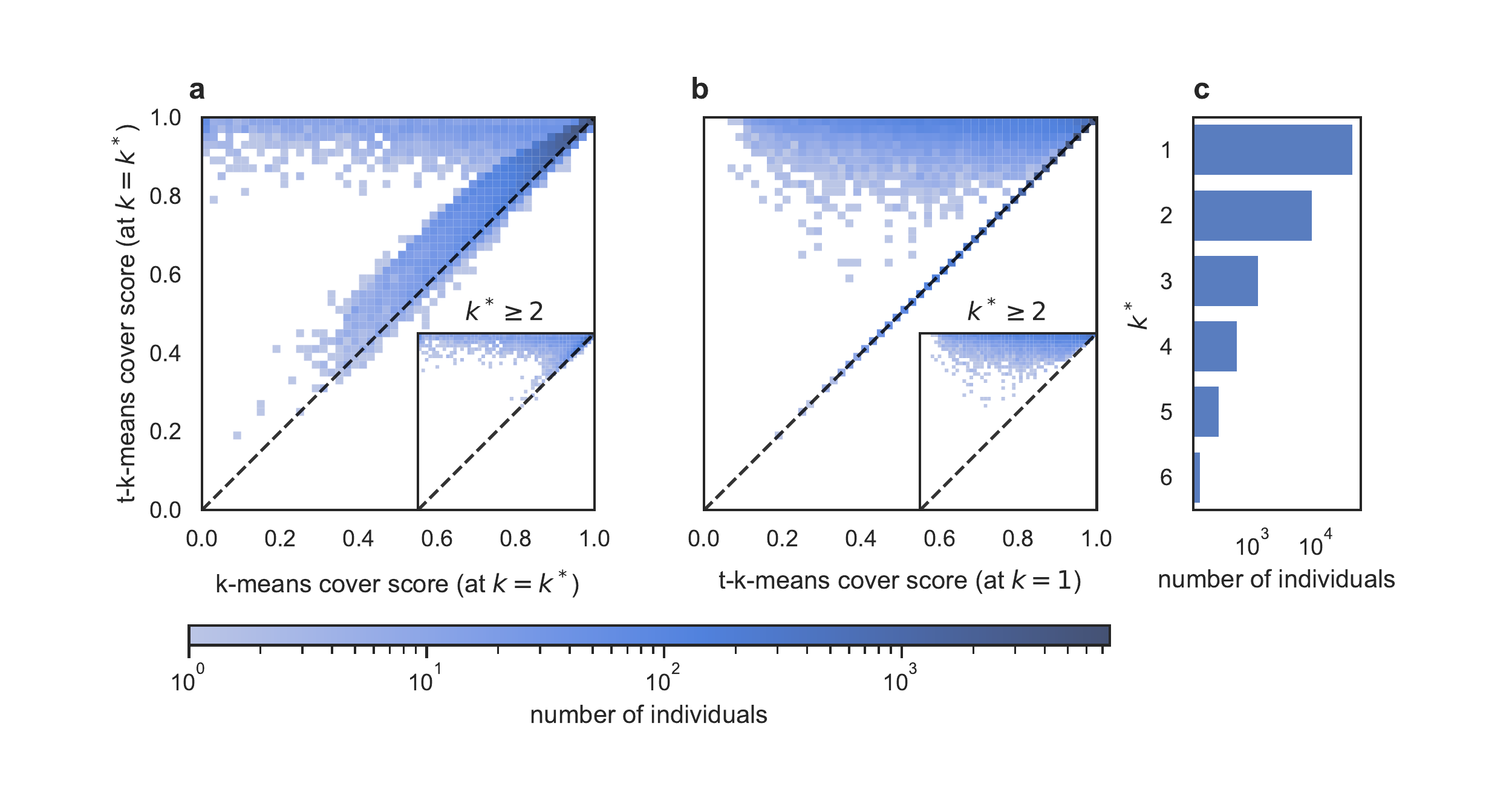}
    \caption{\textbf{Polycentric activity spaces in empirical data.} \textbf{(a) t-k-means outperforms k-means.} Distribution of the cover score for centres obtained using the t-k-means vs. the k-means methods for $k=k^*$. The inset depicts the same distribution restricted to polycentric individuals. 
    \textbf{(b) The flexicentric description outperforms the monocentric description.} Distribution of cover score of centres obtained setting $k=k^*$ vs $k=1$ using the t-k-means algorithm. Elements on the diagonal correspond to the case when $k^*=1$.
    The inset depicts the same distribution restricted to polycentric individuals. \textbf{(c) Distribution of the optimal number of centers $k^*$.} Results are shown for dataset D2 (see Supplementary Information Fig.~\ref{fig:empirical_results_D1} for dataset D1).
    }
    \label{fig:empirical_results_D2}
\end{figure}

Importantly, using empirical data, we find that the centres identified by t-k-means achieve a higher cover score compared to those identified by the simple k-means algorithm (see Fig.~\ref{fig:empirical_results_D2} and Supplementary Information Fig.~\ref{fig:empirical_results_D1}).

\textbf{Choosing the number of centres.} The cover score can be used to compare the quality of the solutions obtained by the t-k-means algorithm for different values of $k$. We note that increasing the number of centres increases the complexity of the model. Thus, we expect the quality of description to increase as well. In the extreme situation when $k = |L|$, the cover score is equal to $1$ (see Supplementary Information for details).

Since the problem we face is similar to the problem of choosing the value of $k$ in the k-means clustering, we adopt a widely used solution to the latter, the \emph{gap statistic} method \cite{tibshirani2001estimating}. This method allows us to evaluate if the increase in cover score for larger values of $k$ is significantly higher than expected simply due to the increase in model complexity. The method works as follows. Let $x_k$ be the cover score of the clustering with $k$ centres. To provide a baseline, we generate $100$ 'reference' datasets with the same set of weights, each by sampling uniformly within the bounding box of the original data. Let $y_k$ and $sd_k$ be the mean and standard deviation respectively of these reference datasets' cover scores. We define the gap statistic as $gap(k) = x_k - y_k$. To find the optimal value of $k$, we first set an upper bound (6 in our simulations), and choose the maximum value of $k^{*}$ such that its gap statistic is a significant improvement over the highest gap statistic of $k < k^*$. That is, $k^*$ is the maximum value of $k$ such that $gap(k^*) - sd_{k^*} > gap(k')$, where $k' = \argmax_{k<k^*}gap(k)$. Using synthetic data, we verified that the gap-statistic method is reliable when clusters are sufficiently disjoint, and is also sensitive to outliers (see Supplementary Information Fig.~\ref{fig:synth_2_normal}).

\textbf{Empirical Results.} We apply the method described above to the trajectories of $\sim 50,000$ individuals from two datasets: $D1$ describes the movements of $566$ students at the Technical University of Denmark, and $D2$ describes the movements of $\sim 49,000$ individuals from Germany, Japan and the United Kingdom (see further details in the \emph{Methods} section). Using the gap statistic method, we find that $42.8\%$ and $22.3\%$ of individuals are polycentric ($k^* \geq 2$) in D1 and D2 respectively (see Supplementary Information Table~\ref{tab:k*_data}). Further, the centres identified by t-k-means have larger cover score than those identified by the standard k-means algorithm: t-k-means strictly outperforms k-means for 43921 (90.5\%) of the individuals, including 8516 (78.6\%) out of 10828 polycentric individuals (see Fig.~\ref{fig:empirical_results_D2} for results on D2, and Supplementary Information Fig.~\ref{fig:empirical_results_D1} for results on D1).

\textbf{Validation with empirical data.} Given a set of locations with corresponding weights, the algorithm described above identifies the number and position of centres of activity. But does the flexicentric activity space description capture individual whereabouts without overfitting? 
We validate our approach by comparing the ability of the monocentric and the flexicentric descriptions, as estimated using only a fraction of an individual's mobility trace, to describe unseen data points. 

In line with previous literature \cite{fleming2017new,downs2008effects}, we use a simple Gaussian mixture to model how individuals allocate time in the activity space.
In particular, we propose that, given an activity space characterised by a set of $k$ centres $\{ \mathbf{c}_1, \mathbf{c}_2, \dots \mathbf{c}_k \}$ and radius of gyration $r_g$, the occupation probability $P(x)$, characterising the likelihood of finding the individual in a given location $x$ is described by a superposition of $k$ 2-dimensional symmetric Gaussians. 
Each Gaussian is centred in one of the centres and has equal variance of $\sigma^2=(\frac{r_g}{\sqrt{k}})^2$.
To get an accurate description, we propose that each Gaussian distribution has weight equal to the total weight of the locations associated with its centre. 
Thus, we have that the occupation probability is: 
\begin{equation}
 P(x) = \sum_{\alpha} W_{\alpha}\mathcal{N}(\mathbf{c}_\alpha,\,\sigma^{2})(x)\
\label{eq:time_allocation}
\end{equation}
, where $\mathcal{N}(\mathbf{c}_\alpha,\,\sigma^{2})$ is a Gaussian distribution with parameters $\mathbf{c}_{\alpha}$ and $\sigma^{2}$, and $W_{\alpha}$ is the weight associated to centre $\mathbf{c}_{\alpha}$. Note that the model reduces to a simple monocentric Gaussian model for $k=1$.

Using dataset D1, for each individual, we select a random sample consisting of $80\%$ of all records (the train set), to find: (i) the centre of mass and radius of gyration characterising the monocentric description (see Eq.~\ref{eq:com} and Eq.~\ref{eq:rog}); (ii) the centres (with associated weights) characterising our flexicentric description using the t-k-means algorithm. In this case, for consistency across replicates of the routine, we set the number of cluster $k=k^*$ as found by the gap-statistic method using the entire dataset. By doing a train-test split on each individual 20 times, we verified that in $(91.7 \pm 17.5)\%$ of the cases, $k^*$ of the train set matches that of the entire dataset.
We use the values found in (i) and (ii) as input parameters to the occupation probability in Eq.~\ref{eq:time_allocation} for the monocentric and flexicentric model, respectively. Note that, for individuals identified as monocentric ($k^*=1$) by our method, the two competing models differ only in the position of the centre of the Gaussian, because the t-k-means algorithm does not necessarily identify the centre of mass as the centre.

We evaluate the goodness of the two models at the task of predicting individuals positions in space by computing the log-likelihood \cite{clauset2009power} of the remaining $20\%$ of the data (the test set) under the two models. We repeat this procedure $20$ times for each individual, and use the Welch's t-test \cite{welch1947generalization} (at significance value $\alpha=0.01$) to compare the log-likelihoods of the two competing predictions. We find that, out of $242$ cases of individuals with $k^*>1$, the flexicentric model performs better than the monocentric model in $230$ cases. For $323$ out of $324$ individuals with $k^*=1$, the test is not conclusive, while in one case the monocentric model performs better (see Supplementary Table~\ref{tab:validation_D1} and Fig.~\ref{fig:validation_D1}). To check that our approach is not overfitting, we compare the log-likelihoods of the train and test data under the flexicentric model, and find that they differ significantly only for 37 out of 566 individuals (see also Fig.~\ref{fig:validation_D1}b). As a further validation against overfitting, we compare the Shannon entropy describing the allocation of weight across clusters $\left(-\sum_\alpha W_\alpha \log W_\alpha \right)$ in the train and test sets. We find that the average ratio between the test and train entropies across polycentric ($k^* \geq 2$) users is $0.99 \pm 0.05$, suggesting that the allocation of time is relatively similar in the two sets (see also Supplementary Information Fig.~\ref{fig:cluster_weights_entropy}).
Altogether, the results above reveal that, for individuals with polycentric activity, the flexicentric model captures well individual whereabouts without overfitting.

\begin{figure}
    \centering
    \includegraphics[width=\textwidth]{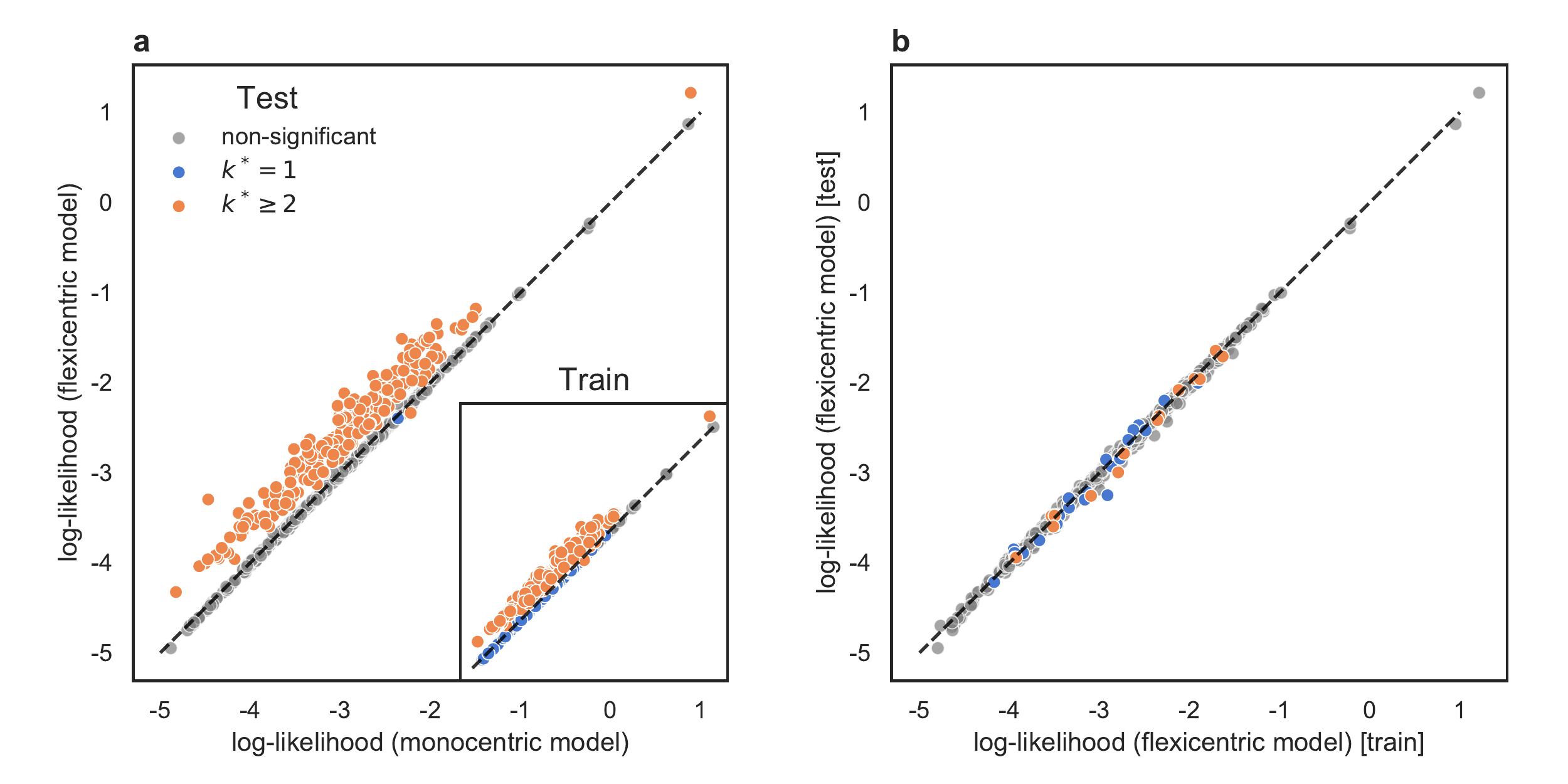}
    \caption{\textbf{Polycentric activity spaces describe individuals whereabouts without overfitting.} \textbf{(a) Log-likelihood of the flexicentric model vs the monocentric model.} Results are shown for data in the test set (larger subplot) and train set (inset).
    \textbf{(b) Log-likelihood of the data in the test vs the train set under the flexicentric model.} In both subplots, only points such that the difference in the two log-likelihoods is significant under Welch's t-test (significance level $\alpha =0.01$) are depicted in colour (monocentric individuals in blue, and polycentric in orange). The rest are plotted in grey.}
    \label{fig:validation_D1}
\end{figure}

\textbf{Understanding heterogeneities of individuals' activity spaces.} Having identified that there is wide heterogeneity across individuals with respect to the number of activity centres (see Supplementary Information Table~\ref{tab:k*_data}), it is natural to wonder which aspects are related to polycentric mobility. In this section, we use dataset D2 to study the relation between the number of centres and individuals socio-demographic attributes, focusing on gender, age, country of origin, and urbanisation level, where the latter is estimated based on individuals home-locations (see the \emph{Methods} section). 

First, we focus on gender alone. 
We find that there is a statistically significant difference between females and males, with $\sim 24.2\%$ of males being polycentric, compared to only $19.1\%$ of females ($p\sim10^{-38}$ under Fisher exact test \cite{raymond1995exact}). 
This result reveals that gender and polycentricity are correlated, and it is robust across the three countries under study (see Table~\ref{tab:countrywise}).

Next, we turn to urbanisation, where, for simplicity, we group individuals into \emph{urban} and \emph{non-urban} (see the \emph{Methods} section). We find no significant difference between the two groups, with polycentric individuals representing $\sim22.7\%$ of non-urban, and $\sim22.2\%$ of urban individuals, respectively ($p=0.212$ under the Fisher exact test\cite{raymond1995exact}). The result reveals that level of urbanisation and polycentricity are not correlated, and it is verified in each of the three countries under study, excluding Germany, where the correlation between the two is significant (see Table~\ref{tab:countrywise}).

To further investigate the correlation between gender and polycentricity, we control for age. We divide the population into three age-groups: 18 to 26, 27 to 50, and 51+ years old. We find that in UK and Germany, the correlation between polycentricity and gender is significant only in the middle age group (27 to 50 years old). Instead, in Japan there is a strong correlation for all age groups (see Table~\ref{tab:countrwise_age}).

\begin{table}[]
    \centering
    \begin{tabular}{|c||cc|c||cc|c|}
    \hline
        \textbf{country} &  \multicolumn{2}{c|}{\textbf{\% polycentric}} & \textbf{p-value} & \multicolumn{2}{c|}{\textbf{\% polycentric}} & \textbf{p-value} \\
        & \textbf{males} & \textbf{females} & & \textbf{urban} & \textbf{non-urban} &  \\
    \hline
    Japan & 24.4 & 19.2 & 6.46e-30** & 22.5 & 22.9 & 0.555\\
    UK & 22.6 & 19.2 & 5.31e-4** & 20.8 & 21.7 & 0.369\\
    Germany & 24.4 & 18.8 & 3.31e-7** & 20.1 & 23.1 & 0.00628** \\
    \hline
    Overall & 24.2 & 19.1 & 1.78e-38** & 22.2 & 22.7 & 0.212\\
    \hline
    \end{tabular}
    \caption{\textbf{Correlation between urbanisation level, gender and the number of centres.} Percentage of polycentric individuals by country and gender (left table) or by country and urbanisation-level (right table). P-values smaller than 0.01 (marked with **) indicate that there is a significant correlation between gender (left) or urbanisation (right) and polycentricity, under the Fisher exact test \cite{raymond1995exact}.}
    \label{tab:countrywise}
\end{table}

\begin{table}[]
    \centering
    \begin{tabular}{|c|c|cc|c|}
    \hline
        \textbf{country} & \textbf{age group} &  \multicolumn{2}{c|}{\textbf{\% polycentric}} & \textbf{p-value}\\
        && \textbf{males} & \textbf{females} & \\
    \hline
    Japan & 18-26 & 23.7 & 18.5 & 9.01e-4**\\
        & 27-50 & 23.7 & 18.9 & 1.68e-16**\\
        & 51 + & 25.9 & 19.9 & 1.15e-11**\\
        &&&&\\
    UK & 18-26 & 19.9 & 18.6 & 0.566 \\
        & 27-50 & 22.6 & 18.8 & 0.00257**\\
        & 51 + & 24.8 & 20.4 & 0.0888 \\
        &&&&\\
    Germany & 18-26 & 25.3 & 19.2 & 0.0139 \\
        & 27-50 & 23.6 & 18.2 & 8.69e-5** \\
        & 51 + & 26.4 & 21.3 & 0.0815 \\
    \hline
    \end{tabular}
    \caption{\textbf{Correlation between age and number of centres.} Percentage of polycentric individuals by country, age group and gender. P-values smaller than 0.01 (marked with **) indicate that there is a significant correlation between gender and polycentricity (given country and age group), under the Fisher exact test \cite{raymond1995exact}.}
    \label{tab:countrwise_age}
\end{table}

\section*{Discussion}
The widely popular radius of gyration  \cite{gonzalez2008understanding} for quantifying activity spaces has recently been challenged by new empirical findings showing that locations visited by individuals are organised in multiple clusters \cite{bagrow2012mesoscopic,alessandretti2020scales}. 
In this paper, we have proposed a generalisation of the radius of gyration, the \textit{flexicentric model}, that properly accounts for the polycentric nature of human activities. 
We have also introduced the \emph{cover score}, a new metric to assess the quality of a polycentric model describing individual whereabouts, and presented a new algorithm to compute individual numbers of centres and their locations. 
Analysing two real-world large-scale datasets of GPS trajectories, we have found that around $42.8\%$ and $22.3\%$ of individuals in our two datasets respectively have polycentric mobility behaviour.
We showed that, for these individuals, the flexicentric model trained on a subset of the entire trajectory allows to better describe unseen data in comparison to the monocentric model, ensuring that it does not overfit.
Interestingly, we discovered that having more than one centre is significantly more likely in men compared to women, especially in the middle age group, across the different countries under study. 
This result is in line with similar empirical findings on the existence of gender gaps in human mobility \cite{gauvin2020gender,alessandretti2020scales}. 
A potential reason why the gender gap is more pronounced in the middle age-group concerns the fact that employed individuals are likely to have at least two centres, if their home and work locations are far enough \cite{bagrow2012mesoscopic}.
The gender gap in employment may thus partly explained the differences in travel behaviour across genders \cite{ng2018understanding}.
In contrast, we found that the level of urbanisation around an individual's home location does not relate to the number of centres. These observations would not have been possible from a monocentric point of view. 
By opening the possibility to characterise individuals whereabouts in a polycentric fashion, our work allows to enrich the description of human mobility, going from a single quantity, i.e. the radius of gyration, to multiple numbers, or spatial scales, associated to distances between centres and around the centres. This new lens to study individual-level travel behaviour may help to improve our quantification of heterogeneities across individuals, which is critical to design accurate mobility models, and ultimately improve our ability to forecast the spread of epidemics and plan transportation that account for the diverse needs of all individuals. In addition to these potential applications, future works include the possibility to study the evolution of individuals whereabouts over time and investigate human mobility in more general metric spaces, where the distance between locations is not purely geographical, but may be built on commuting time information or distance in transport networks.

\clearpage

\begin{small}

\section*{Methods} 

\subsection*{Data Description}
Our study is based on two datasets. 

\textbf{Dataset D1} was collected as part of an experiment that took place between September 2013 and September 2015\cite{stopczynski2014measuring}. The experiment involved $851$ Technical University of Denmark students (about 22\% female and about 78\% male), typically aged between $19$ and $21$ years old. Participants’ GPS records over time was sampled every 5 minutes. Data collection was approved by the Danish Data Protection Agency. All participants provided informed consent. We selected 5 months of data for $566$ individuals from the entire dataset (see Supplementary Information Section S1).

\textbf{Dataset D2} contains anonymised GPS location data collected by a global smartphone and electronics company between 2017 and 2019. The data consist of anonymised users with self-reported age, gender, height, weight and country of residence. Data were extracted through a smartphone app. All data analysis was carried out in accordance with the European Union’s General Data Protection Regulation 2016/679 (GDPR) and the regulations set out by the Danish Data Protection Agency. Individuals provided informed consent to the analysis of their data. We estimated the level of urbanisation and country for each individual from the position of the top-location. Data on the urbanisation level in the area surrounding individuals’ home locations is based on the GHS Settlement Model grid that delineates and classifies settlement typologies via a logic of population size, population and built-up area densities. This classification categorises areas in urban areas, towns and rural areas. In our analysis, we merged towns and rural areas into a single category. Data can be downloaded from: \href{https://ghsl.jrc.ec.europa.eu/data.php.}{https://ghsl.jrc.ec.europa.eu/data.php}. We selected approximately 50,000 individuals from the top three countries in the dataset (Japan, Germany and the UK) with at least six months of data, whose position is known, every day, on average 75\% of the time. Supplementary Tables~\ref{tab:urb_distr} and \ref{tab:gender_distribution} summarise the distributions of urbanisation and gender. We categorise individuals into three age groups: $18 - 26$ years old, $27 - 50$ years old, and $\geq 51$ years old. The summary of the age distribution is included in Supplementary Information Fig.~\ref{fig:age_distribution}.

\subsection*{Data Availability} 

Derived data that support the findings of this study are available in DTU Data with the identifier \href{[AVAILABLE UPON PUBLICATION]}{[AVAILABLE UPON PUBLICATION]}. Additional data related to this paper may be requested from the authors. Raw data for dataset D1 are not publicly available due to privacy considerations, but are available to researchers who meet the criteria for access to confidential data, sign a confidentiality agreement and agree to work under supervision in Copenhagen. Please direct your queries to the corresponding author. Raw data for dataset D2 are not publicly available to preserve individuals’ privacy under the European General Data Protection Regulation.

\subsection*{Code Availability}

Code is available at \href{https://github.com/rohit-sahasrabuddhe/polycentric-mobility.git}{https://github.com/rohit-sahasrabuddhe/polycentric-mobility.git}.

\subsection*{The trimmed k-means algorithm} The traditional k-means algorithm takes as input a set of (possibly weighted) points embedded in a metric space and the number of clusters $k$ \cite{likas2003global}. It iteratively partitions the data points into $k$ clusters as follows. The algorithm is initialised by choosing $k$ centres randomly amongst the data points. In every iteration, points are assigned to their closest centre, and each centre is updated to the (weighted) mean of the points assigned to it. The algorithm terminates when the centres become stationary or when the number of iterations exceeds a pre-defined limit. It is well-known that k-means clustering is sensitive to the initial choice of centres, as well as to outliers in the data. Two types of strategies to deal with these issues are (a) choosing a better initialisation of centres (such as k-means++ \cite{arthur2006k}), and (b) running the clustering several times with different initial centres and choosing the best result \cite{likas2003global}.

The problem of outliers affects location data, because, due to the effects of exploration and preferential return \cite{song2010limits}, the distribution of time spent per location is highly skewed. In order to ensure that places where individuals spend little time do not affect dramatically the solution of the clustering, and inspired by previous modifications of the k-means algorithm \cite{cuesta1997trimmed}, we propose the trimmed k-means algorithm, \emph{t-k-means}. At each iteration of t-k-means, the centre is updated to the (weighted) mean of points corresponding to $t$ fraction of weight closest to the centre. This reduces the effect of points that are assigned to the cluster, but are far from the centre. We note that setting the trimming parameter $t = 1$ corresponds to running the traditional k-means algorithm. The value of $t$ is set to $0.9$ for the results presented in this paper, and are robust to variations of this parameter (see Supplementary Information Fig.~\ref{fig:trimming_parameter_sensitivity}). Further, we deal with the sensitivity to initialisation by clustering each dataset $50$ times, and then picking the result that minimises the inertia.

\end{small}

\textbf{Acknowledgements} RS is supported by the DST-Inspire scholarship (Government of India).\\
\textbf{Author Contributions.} RS, RL and LA designed the study.  RS performed the analyses and implemented the code. All authors analysed the results and wrote the paper. \\
\textbf{Competing Interests.} The authors declare that they have no competing financial interests.\\


\pagebreak
\begin{center}
\textbf{\LARGE Supplementary Information \\ 
From centre to centres: polycentric structures in individual mobility}
\end{center}

\setcounter{equation}{0}
\setcounter{figure}{0}
\setcounter{table}{0}
\setcounter{page}{1}
\setcounter{section}{0}

\makeatletter
\renewcommand{\thesection}{S\arabic{section}}
\renewcommand{\theequation}{S\arabic{equation}}
\renewcommand{\thefigure}{S\arabic{figure}}
\renewcommand{\thetable}{S\arabic{table}}



\section{Data description and pre-processing}

For both datasets, the raw data of mobility traces is first put through the Infostop algorithm \cite{aslak2020infostop} with parameters $r_1,r_2=30$ metres, $t_{min} = 10$ mins. This gives us, for each individual, a set of records, each corresponding to the individual staying within a 30 meter radius for at least 10 minutes. For each record, we also have the latitude and longitude, and the amount of time spent at that stop location. 

\textbf{Filtering of dataset D1.} With this data, we perform the following steps to clean the dataset D1 and choose a subset to analyse:
\begin{itemize}
    \item We first discard all records with time-spent $\geq$ 7 days.
    \item We restrict the analysis to locations in the Sjælland and Hovedstaden administrative regions (as per the shapefile obtained from https://diva-gis.org/ accessed 20-Aug-2020). This region contains 88.8\% of records corresponding to 91.6\% of the time spent.
    \item We then consider records from the Spring semester (January - May) of 2014 which corresponds to 151 days. For each individual, we calculate the active time as the total amount of time for which we know their stop location. We note that the distribution of active times is bimodal, and thus restrict our dataset to individuals of high active time by placing a cutoff at the 68 day mark.
    \item At the end, we arrive at a dataset of 566 individuals, each with more than 68 days worth of data in the region of interest.
\end{itemize}

\textbf{Filtering of dataset D2.} We perform the following steps to clean and filter the dataset D2:
\begin{itemize}
    \item We discard all records with time-spent $\geq$ 7 days.
    \item We select users whose average daily time-coverage (fraction of time an individual's location is known) larger than $75\%$
    \item We select individuals whose country of origin is either Japan or Germany or the UK. These are the top three countries by number of users in the entire dataset. 
    \item For each individual, we select the first $180$ days of data.

\end{itemize}

\begin{figure}
    \centering
    \includegraphics[width=0.7\textwidth]{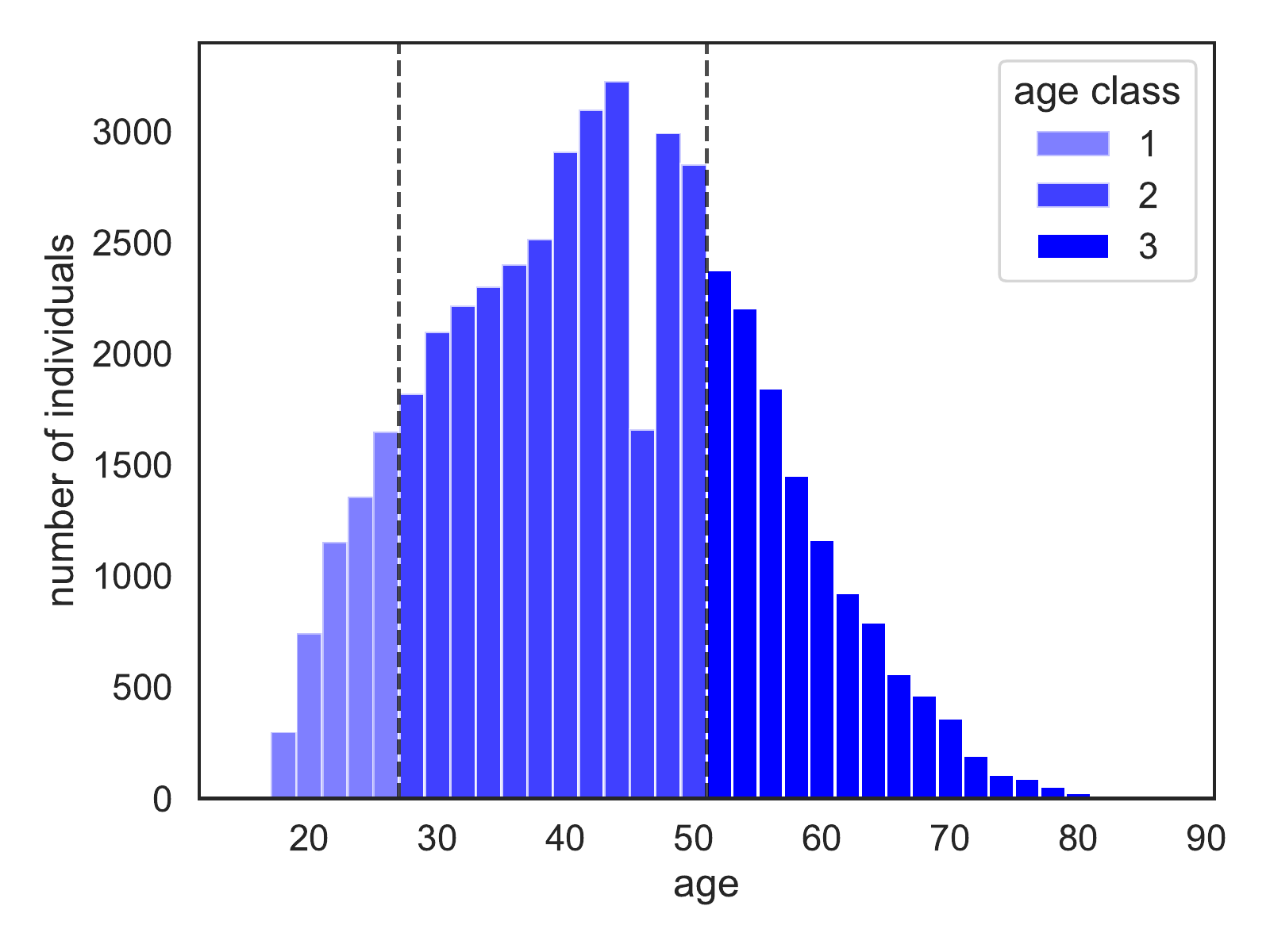}
    \caption{Age distribution, and classification into age groups of individuals in dataset D2.}
    \label{fig:age_distribution}
\end{figure}

For the users selected in dataset D2, we present the distribution of age in Fig.~\ref{fig:age_distribution}, the distribution of urbanization level in Table~\ref{tab:urb_distr} and the distribution of gender in Table~\ref{tab:gender_distribution}.

\begin{table}[h!]
    \centering
    
    \begin{tabular}{|c|c|c|c|c|}
        \hline
        \textbf{Country} & \textbf{Rural (1)} & \textbf{Semi-urban (2)} & \textbf{Urban (3)} & \textbf{Total}\\
        \hline
         Japan & 1903 & 5061 & 28644 & 35608\\
         UK & 918 & 1822 & 4289 & 7029 \\
         Germany & 1474 & 2075 & 2348 & 5897\\
         \hline
         \textbf{Total} & 4295 & 8958 & 35281 & 48534 \\
         \hline
    \end{tabular}\vspace{1cm}
    \caption{Number of individuals per country and urbanization-level in dataset D2.}
    \label{tab:urb_distr}

 \end{table}
 \begin{table}[h!]
   \centering
    \begin{tabular}{|c|c|c|c|}
        \hline
        \textbf{Country} & \textbf{Males} & \textbf{Females} & \textbf{Total}\\
        \hline
         Japan & 23187 & 12421 & 35608\\
         UK & 4091 & 2938 & 7029\\
         Germany & 3280 & 2617 & 5897 \\
         \hline
         \textbf{Total} & 30558 & 17976 & 48534 \\
         \hline
    \end{tabular}
    \caption{Number of individuals per country and gender in dataset D2.}
    \label{tab:gender_distribution}
\end{table}

\clearpage

\section{The t-k-means algorithm}
\begin{figure}[h]
    \centering
    \includegraphics[width=\textwidth]{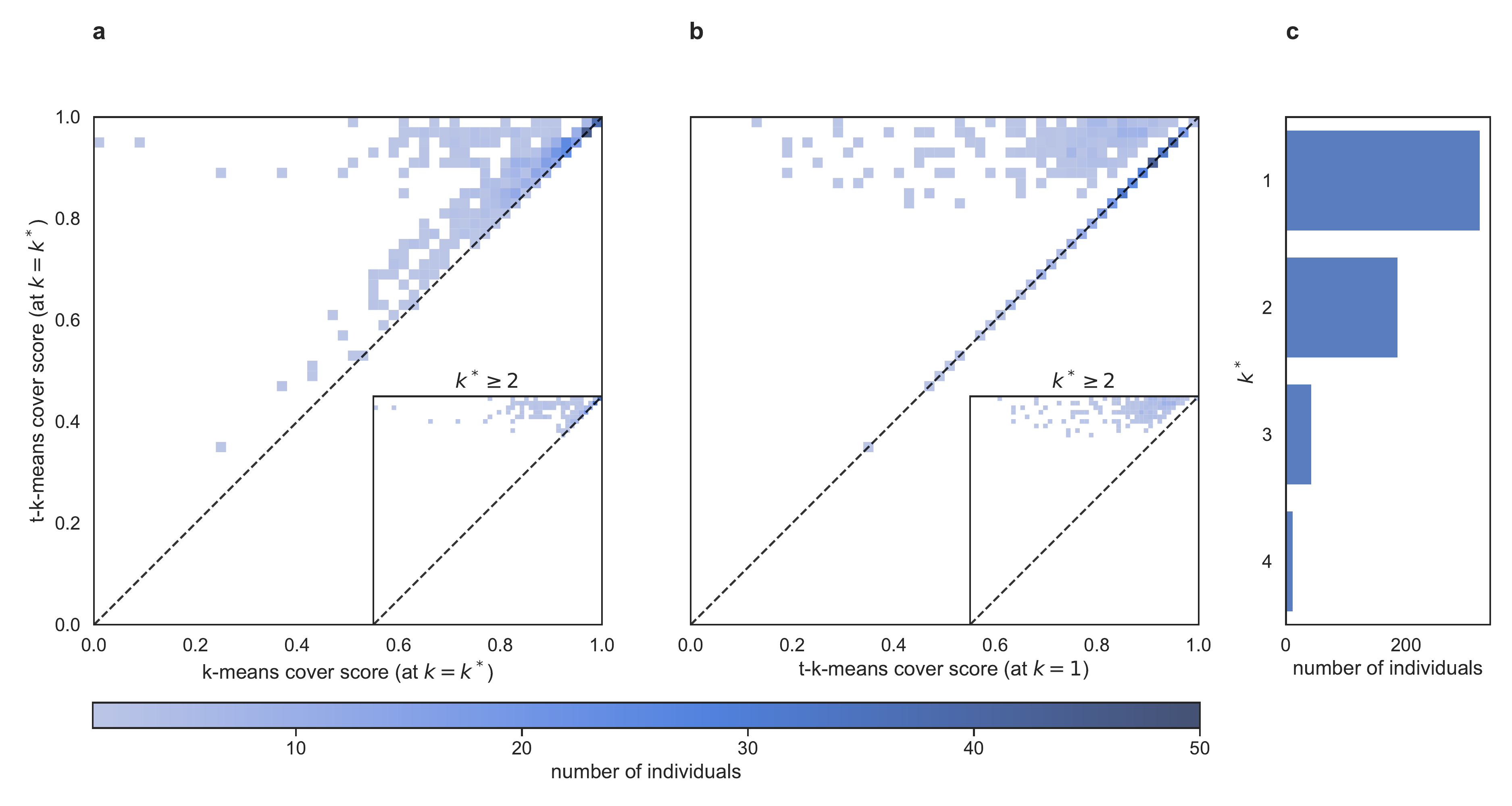}
    \caption{\textbf{Results on D1.} \textbf{(a) t-k-means outperforms k-means} Distribution of the cover score for centres obtained using the t-k-means vs. the k-means methods for $k=k^*$. The inset depicts the same distribution restricted to polycentric individuals. 
    \textbf{(b) The flexicentric description outperforms the monocentric description.} Distribution of cover score of centres obtained setting $k=k^*$ vs $k=1$ using the t-k-means algorithm. The inset depicts the same distribution restricted to polycentric individuals.
    \textbf{(c) Distribution of $k^*$.}}
    \label{fig:empirical_results_D1}
\end{figure}

Empirical results on applying the t-k-means algorithm on dataset D1 are presented in Fig.~\ref{fig:empirical_results_D1}.\\
\textbf{Considerations on the relation between the cover score and the number of centres.}
Increasing the number of centres $k$ increases the complexity of the model, and thus we expect an increase in the quality of description. However, we note that the cover score can either increase or decrease when increasing $k$. For example, creating a new centre in a sparse region could result in the cover score decreasing, because (a) the area around the other centres is reduced, and (b) the sparse area around the new centre is given more importance. In this respect, there is a subtle but important difference with respect to the k-means clustering algorithm, in which the inertia is used (see Eq.~\ref{eq:inertia}) to assess the quality of a solution. The inertia, in fact, decreases monotonically with increasing $k$, because adding a centre results necessarily in a better description.
\begin{table}
    \centering
    \begin{tabular}{|c|cc|}
    \hline
        \textbf{$k^*$} &  \multicolumn{2}{c|}{\textbf{number of individuals}} \\
        & \textbf{D1} & \textbf{D2}\\
    \hline
        1 & 324 &37706\\
        2 & 187 & 8634 \\
        3 & 43 & 1206\\
        4 & 12 & 555\\
        5 & 0 & 287\\
        6 & 0 & 146\\
    \hline
    \end{tabular}
    \caption{\textbf{Distribution of $k^*$ for empirical datasets D1 and D2.}}
    \label{tab:k*_data}
\end{table}

\begin{figure}[h]
    \centering
    \includegraphics[width=0.6\textwidth]{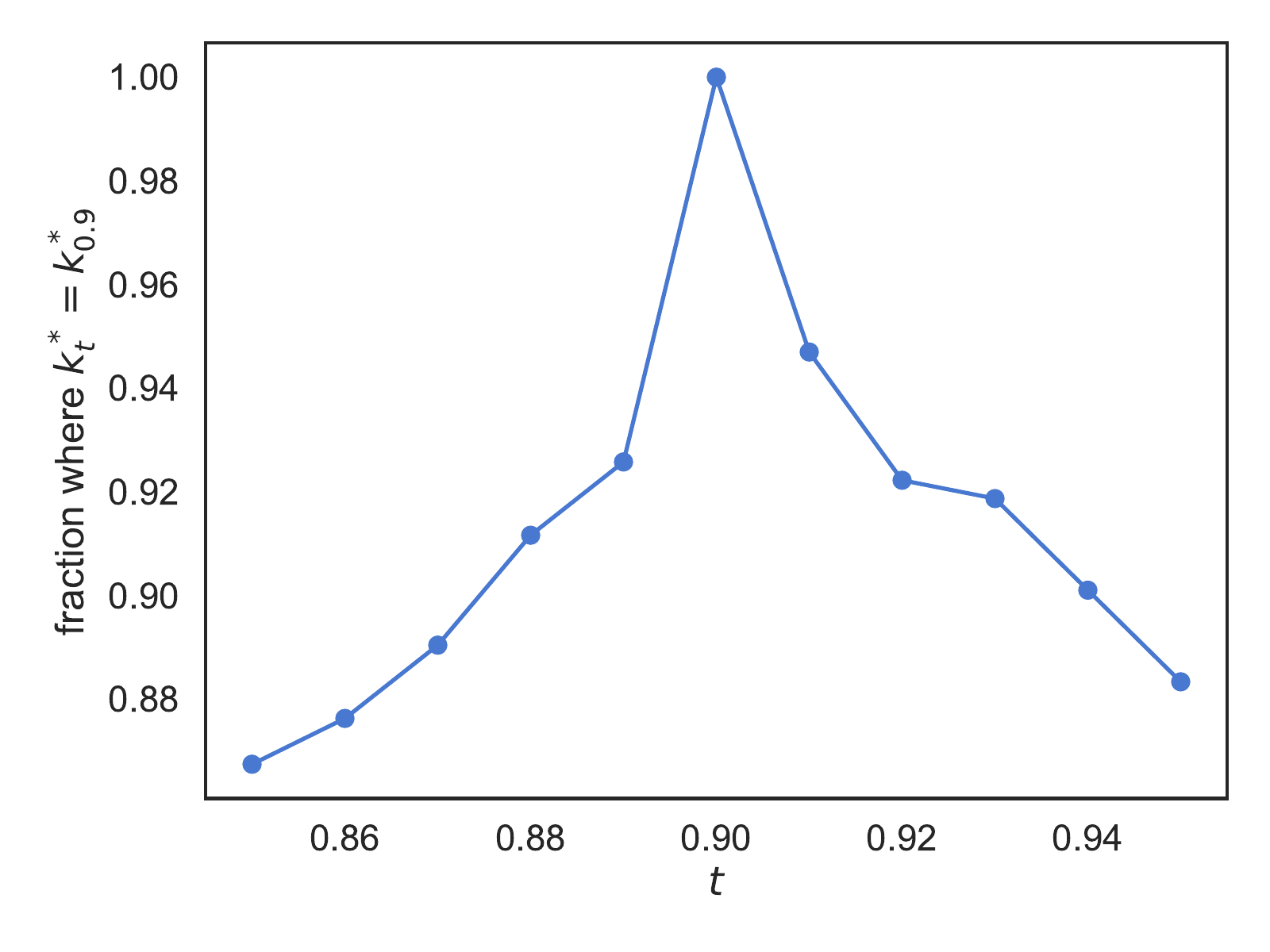}
    \caption{Sensitivity of $k^*$ on the choice of trimming parameter $t$.}
    \label{fig:trimming_parameter_sensitivity}
\end{figure}

\textbf{Sensitivity to changes in trimming parameter $t$.} We check the sensitivity of the results to variation in $t$ by running t-k-means on all individuals in the D1 dataset for $t \in [0.85, 0.95]$, and call the value of $k^*$ as $k^*_t$. Note that in this notation, the results in the main text correspond to $k^*_{0.9}$. In Fig.~\ref{fig:trimming_parameter_sensitivity}, we plot the fraction of individuals for whom the $k^*_t = k^*_{0.9}$.

\clearpage

\section{Validation against synthetic data.} 
\begin{figure}[h!]
    \centering
    \includegraphics[width=\textwidth]{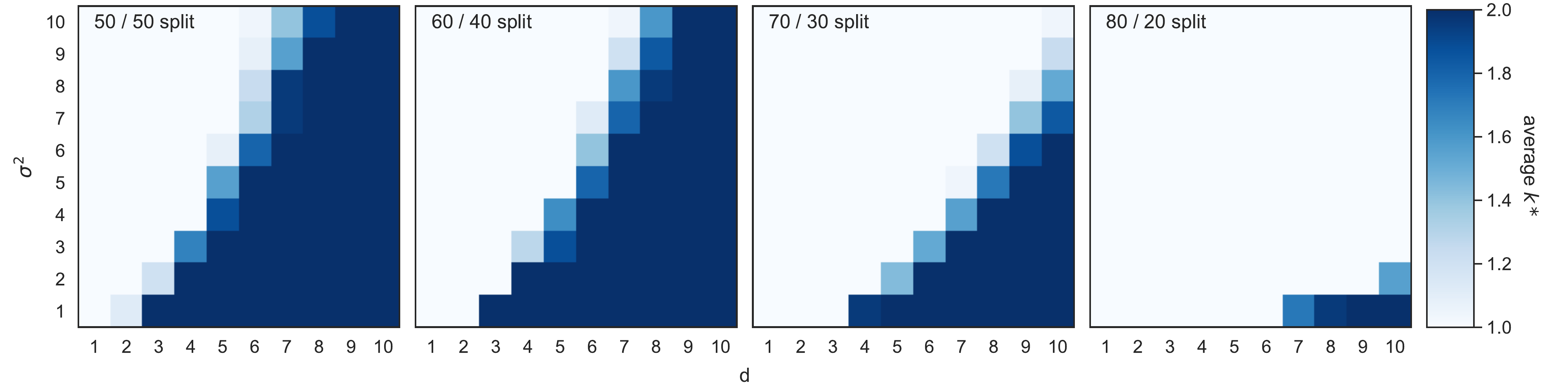}
    \caption{\textbf{Extracting clusters from synthetic data: }The plots depict the value of $k^*$ identified by our algorithm for synthetic data comprising of points sampled from 2 bivariate Gaussian distributions. The 4 panels show (from left to right) the results for a 50/50, 60/40, 70/30, and 80/20 split. Each cell in the heatmap corresponds to a particular value of $\sigma^2$ (variance of each Gaussian along both axes) and $d$ (distance between their centres), and depicts the average value of $k^*$ across 25 runs over randomly generated datasets.}
    \label{fig:synth_2_normal}
\end{figure}

We first validate our approach for selecting $k^*$ using synthetic data. We consider datasets of $100$ locations sampled from two bivariate Gaussian distributions with variance $\sigma^2$ (along both axes) and distance between their centres $d$, implying the number of centre is $k^*=2$. For simplicity, we set all locations to have equal weights, but sample a different number of points from each Gaussian. In particular, we sample points from the 50/50, 60/40, 70/30, and 80/20 splits. In each case, we vary $d$ and $\sigma^2$ in the range $[ 1, 10 ]$, and record the value of $k^*$ across $25$ runs over randomly generated datasets for each pair. We find that the value of $k^*$ identified by the algorithm has desirable properties (Fig \ref{fig:synth_2_normal}). First, for increasingly uneven splits, the algorithm is more likely to predict $k^*=1$, implying the method is robust to outliers. Second, for each split, the algorithm is more likely to predict $k^*=2$ when $d$ is high and $\sigma^2$ is low, which is also desirable. 

\clearpage

\section{Validation against empirical data}
Results for the empirical validation of our method using a tran/test split are presented here and summarized in Table~\ref{tab:validation_D1}. \\

\textbf{Monocentric individuals: }Of the 324 monocentric individuals, we find that the traditional monocentric model outperforms our model for 84 individuals in the train set, but just 1 individual in the test set. Our method never outperforms the traditional method.\\\\
\textbf{Polycentric individuals: }Of the 242 polycentric individuals, our model outperforms the traditional method in 237 and 230 individuals in the train and test set respectively. In turn, the traditional method is significantly better for 2 and 1 individual respectively.\\\\
\textbf{Train vs test: }Comparing the log-likelihoods of our model on the train and test sets shows that their means differ statistically significantly only for 37 individuals out of 566. The log-likelihood of the train set is significantly greater for 25 of these, and that of the test set is greater for the remaining 12.

\begin{table}[h]
    \centering
    \begin{tabular}{|c||cc||cc|}
        \hline
        & \multicolumn{2}{c||}{\textbf{Train set}} & \multicolumn{2}{c|}{\textbf{Test set}}\\
        \hline
        & flexicentric & monocentric & flexicentric & monocentric\\
        \hline
        \textbf{Monocentric} (324) & 0 & 84 & 0 & 1\\
        \textbf{Polycentric} (242) & 237 & 2 & 230 & 1\\
        \hline
    \end{tabular}
    \caption{\textbf{Validation of the flexicentric method on D1.} The values under the 'flexicentric' (resp. 'monocentric') correspond to the number of individuals for whom the log-likelihood of data under the flexicentric (resp. monocentric) model was significantly better. The significance of the difference is established via Welch's t-test, and the values in this table correspond to a p-value $<$ 0.01.}
    \label{tab:validation_D1}
\end{table}

\begin{figure}
    \centering
    \includegraphics{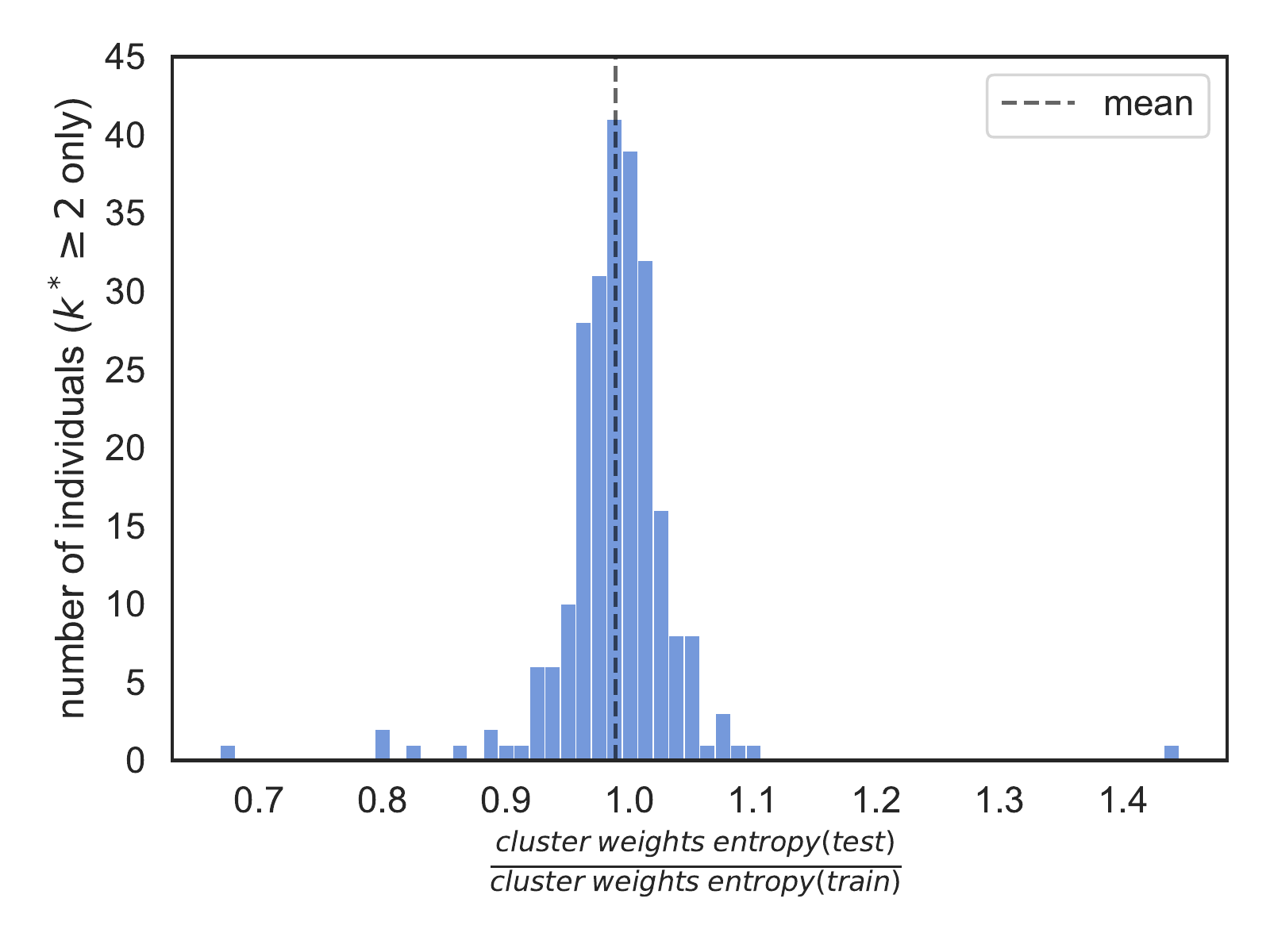}
    \caption{The distribution of the ratio of the Shannon entropy of the weight distribution across clusters in the test and train sets for polycentric users in D1. The mean is 0.99 $\pm$ 0.05.
    }
    \label{fig:cluster_weights_entropy}
\end{figure}

\pagebreak

\bibliographystyle{unsrt}
\bibliography{bibliography}

\end{document}